\documentclass[twocolumn]{aa}
\usepackage{graphicx}
\usepackage{txfonts}

\begin{document}

\title{Research Note: Theoretical Color-Magnitude Diagrams
and the Star Forming Histories of Interacting Open
Multi-population Model Galaxies: Bursts and Busts}

\titlerunning{Observables for Open Multipopulation
Stochastic Galactic Models}

   \author{Giada Valle \inst{1}, Michele Cignoni \inst{1},
Steven N. Shore \inst{1,2}}

\authorrunning{Giada Valle et al.}

\institute{(1) Dipartimento di Fisica ``Enrico Fermi'',
Universit\`a di Pisa, largo Pontecorvo 3, Pisa I-56127 Italy;\\
(2) INFN - Sezione di Pisa, largo Pontecorvo 3, Pisa I-56127, Italy}

   \offprints{Dr. G. Valle}

   \date{accepted 13-06-05}

   \abstract{

This note presents theoretical color-magnitude diagrams (CMDs) and metallicity
evolution for Galactic multipopulation models coupled to stellar evolution
models for systems undergoing stripping and re-accretion of ambient
material, called ``open'' systems in our recent studies (Valle et
al. 2005). We show that the observables for such systems, in
particular those related to the recovered star formation as a function
of time, are ambiguous -- systems with non-monotonic star formation
rates can appear as either bursting systems or galaxies with a hiatus
in the star formation.

   \keywords{Galaxies: theoretical; Star Formation: galactic,
Galaxies: evolution }
   }

   \maketitle

\section{Introduction}

This research note extends the work of Valle, Shore, \& Galli
\cite{valle05} (hereafter VSG05) concerning the effects of environment
on the evolution of galactic populations and metallicities.  Our
intent here is to  underline some observational consequences of this
approach.  To place our simulations in context,  we first summarize
some basic points  already extensively treated in our previous work.
Our {\it standard} model of Galactic chemical evolution uses  a
multi-zone, multi-population approach ({\it cf.} Ferrini et
al. \cite{fmpp92}, Shore \& Ferrini \cite{sf95}, Valle et
al. \cite{valle}).  The system is schematicized using three
distinct zones -- the halo (HA)
(in our models this implicitly includes the
bulge and spheroid), thick disk (TD), and thin disk (DI)
-- that exchange mass, each consisting of three interconverting
phases: diffuse gas ($g$), clouds, ($c$), and stars ($s$). We use a
coupled population dynamical approach which, although without explicit
thermomechanical prescriptions for material transport or
chemodynamical feedback, provides the star formation rate as a
function of time directly from the model equations and all chemical
evolution is followed without instantaneous recycling.

In VSG05 we used this approach to study stochastic accretion and mass
loss within closed and open systems but concentrated on the star
formation rates and chemical evolution.  However, because we have
self-consistent results for both, we can link them to produce
population synthetic color-magnitude diagrams (CMDs) and other observable
diagnostics of galactic evolution.   One of the most interesting
results to emerge from VSG05 was, in our view, the phenomenon of a
{\it burst in reverse} star formation  -- that following a
collision, subsequent replenishment of the interstellar medium through
mass shedding by evolving stars would restart star formation after a
considerable delay, producing the appearance of a bursty star forming
history.  This was notable since there is no provision taken in the
models for enhanced induced star formation during either the
collisional or stripping/refilling events.  This also, under extreme
conditions, led to galaxies with inverted metallicity-age relations,
younger populations that can have lower metallicity than more evolved
components depending on the course of the re-accretion process, a
phenomenon particularly relevant for galaxies in clusters.  For this
reason, we follow up in this paper only simulations of collisions and
collision/stripping/re-accretion scenarios and compare these to the
standard case. We updated the model by including the Kroupa initial
mass function (IMF) (Kroupa \cite{kroupa02}), replacing that obtained
from analysis of molecular cloud fragmentation used in all previous
papers in this series ({\it i.e.} Ferrini et al. \cite{fmpp92}).
All process
rates and chemical yields have been recomputed using the newly adopted
IMF to be consistent with the stellar evolution simulations.

\section{Population Synthesis Code}

The population simulations use the Pisa evolutionary library
(e.g. Cariulo et al. ~\cite{cari04}).  The evolutionary  tracks used
in the simulations were computed assuming a primordial helium
abundance $Y_{P}=0.23$ and a fixed enrichment ratio $\Delta Y/\Delta
Z= 2.5$ (see Pagel \& Portinari 1998;  Castellani et
al. 1999).  In part because of the schematic nature  of the models we
present here, we have not included  a binary population.  For the same
reason we have  not varied the IMF, we still lack detailed knowledge
of the statistical distribution of the mass ratio $q$ and even the
fraction of binaries present in the solar neighborhood.  Because the
evolution of close binaries also depends on the initial period
distributions, the range of phenomenology is far wider than even for
passive (widely separated independently evolving) components. In
addition, because the mass ratios are drawn from what must be a
continuous distribution, binaries will smear the width of the giant
branch, and main sequence, without necessarily producing substructure
(see Hurley \& Tout 1998); in principle, this displacement
can introduce a gap in the RGB region, but it is very
improbable that all binaries have a companion
of equal mass.   Although the Monte Carlo procedure we use here is
designed to model the Solar neighborhood based on the Hipparcos
measurements,  which is dominated by small samples, the same methods
can be used to model a galaxy assuming as inputs the star formation
rate and metallicity-age relation.  There is no gradient in the
metallicity, no structure, and all parameters for the stellar
population are assumed to be independent of position.  Thus, we
emphasize the schematic nature of this calculation - as we did in VSG05
for the star-forming and abundance histories --  and although we model the
full galactic evolution we include only a one zone calculation (thin
disk history) for the CMDs.

An essential difference between our  approach to the simulated field
CMDs and others in the literature has to do with the simultaneous
solutions for the metal abundances and the star formation through the
model equations.  Any population model  depends on two inputs: the
age-metallicity relation and the star formation rate (for fixed IMF)
as a function of time.  In principle, any chemical evolution code
produces a consistent history for the abundances {\it once an assumed
star formation rate is explicitly supplied}.  The systematic bias
introduced by a specific functional choice for the SFR  cannot be
compensated by simply  adjusting parameters.  Instead, different
histories are usually tried and the results compared with
observations.  Since we have a model system for the evolution
equations, at least the systematic bias can be reduced -- although not
eliminated -- because a large variety of interlocking processes are
treated simultaneously.

The simulations were performed using the Monte Carlo
method previously described by Castellani et al. (2002) and Cignoni et
al. (2003); no mass loss is used for the RGB so the horizontal branch
is more prominent in the simulated CMDs than in real galaxies. It
consists of sampling a library of metallicity dependent isochrones
with a constant initial mass function according to the chemical
history.   The full nonlinear galactic model
produces a (numerical) time dependent
star formation rate, $\psi(t)$.  The number of stars  formed before
some time $t_i$, $N(t\le t_i)$, is equated to a random number
and the distribution of times $\{t_i\}$ is then used to sample the
evolutionary tracks with the appropriate metallicity $Z(t_i)$.  The
masses are distributed according to an assumed time-invariant initial
mass function, in this case the same one used for the galactic
evolutionary model.   Each track is  selected based on the
age-metallicity relation for the system derived from the star
formation model, $Z(t)$, and to keep as close
as possible to observables, we used the  color transformations from
Castelli et al. (~\cite{castelli97})  that are also used in the standard
version  of the solar neighborhood population simulation
(e.g. Castellani et al. ~\cite{cast02},  Cignoni et
al. ~\cite{cig03}).

\section{Simulation of Observables}

\begin{figure}
\vspace{5mm}
\centering
\includegraphics[width=8cm]{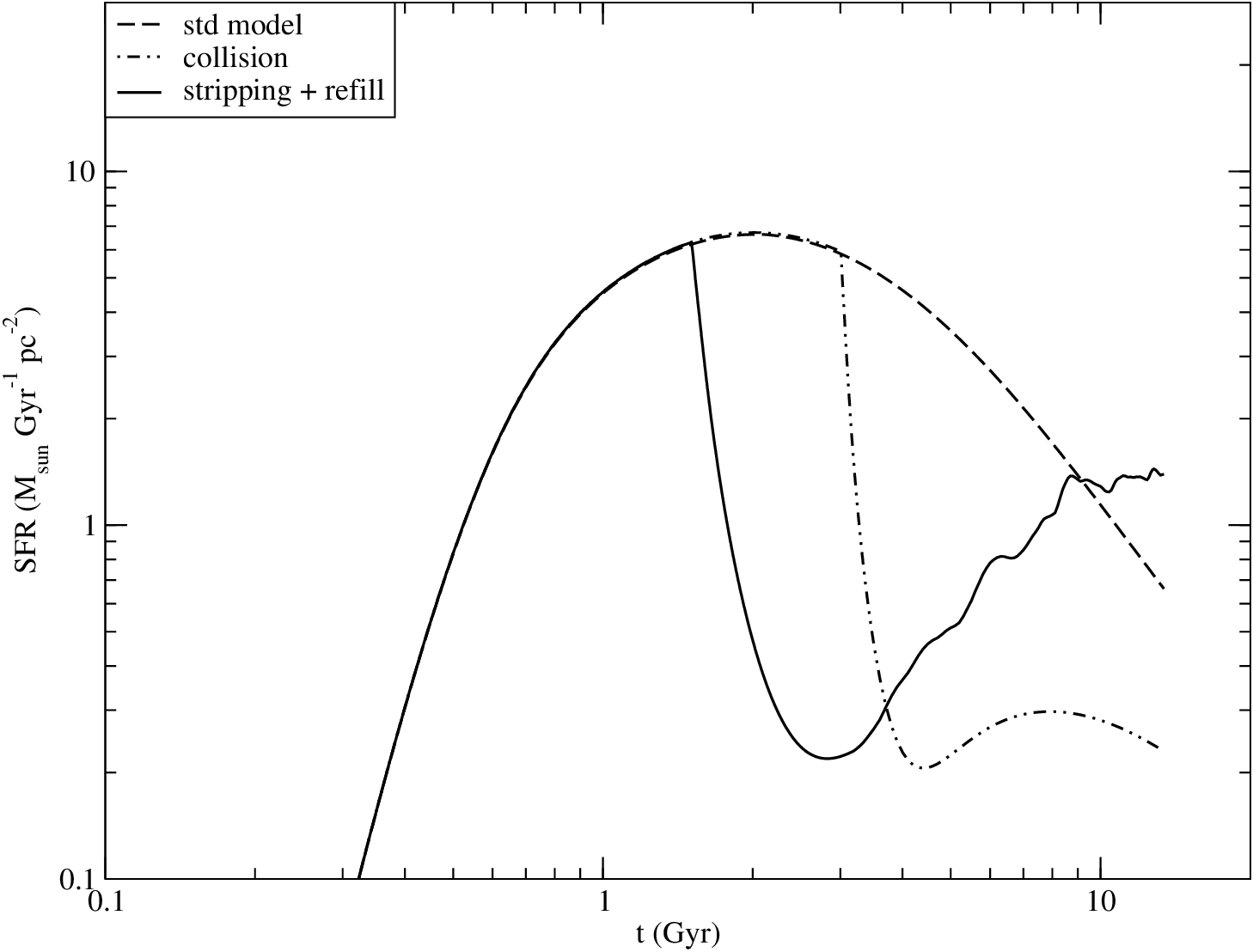}\vspace{15mm}\\
\includegraphics[width=8cm]{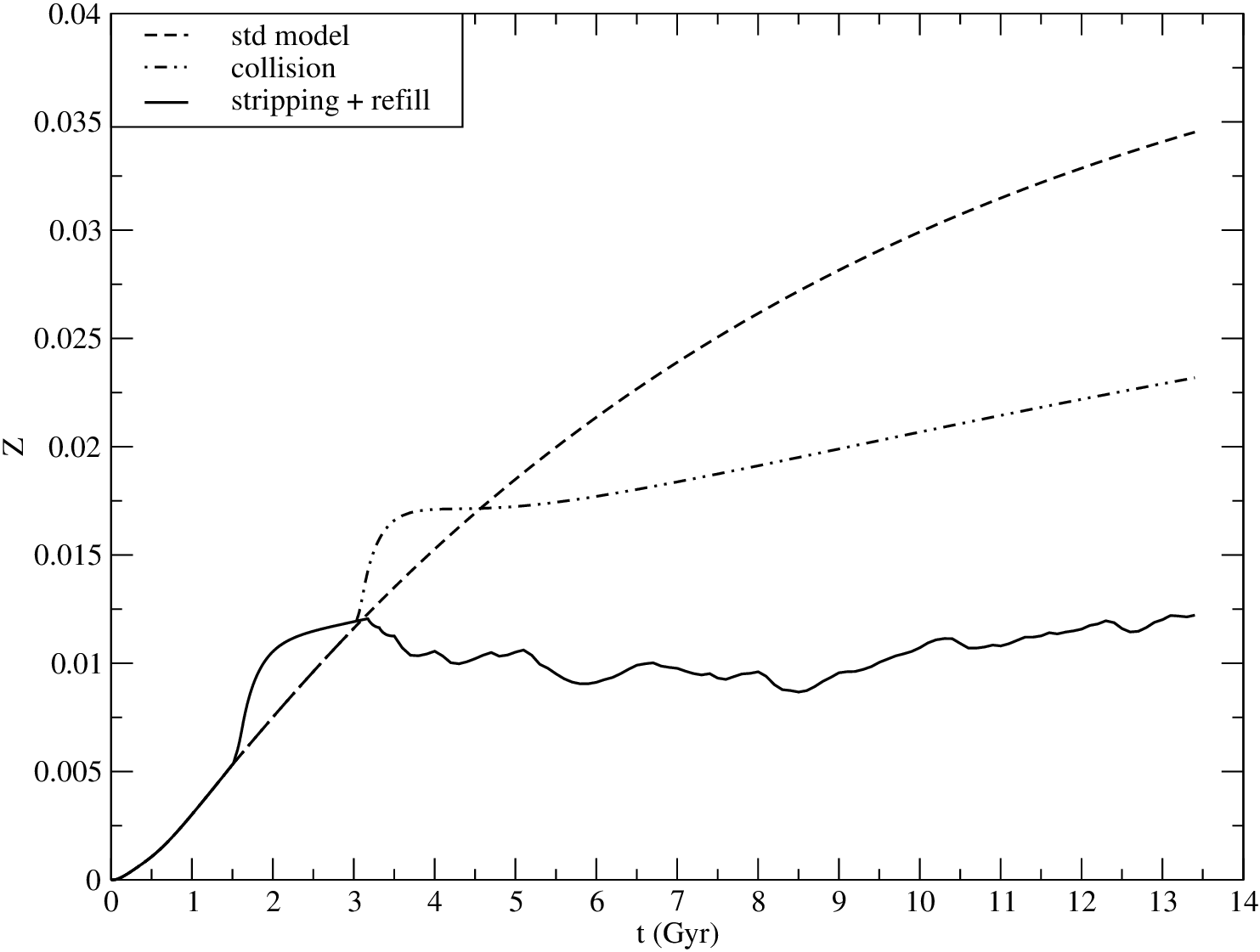}
\caption{{\bf (a)} Star formation rate (top) and {\bf (b)}
 metallicity evolution for the three scenarios (bottom).
{\it Dot-dash}: Collision simulation with $t_{start}=3$ Gyr
and  $\Delta t_{coll}=130$ Myr.  Mass loss profiles were as in Quilis et al.
(\cite{quilis00}).
{\it Black continuous}: Combined effects of collision and environmental infall.
The collision event was assumed to have $t_{start}=1.5$ Gyr
and $\Delta t_{coll}=130$ Myr.
After the collision, the model assumes stochastic infall
and re-accretion of primordial material starting at $t=3$ Gyr.
About 40\% of the
 total initial mass was removed during the collision and then about
 25\% was restored during the subsequent re-accretion.
The comparison with the standard model case is also shown ({\it dash}).
Note: In this figure we display only the thin disk
evolution and update the VSG05 IMF (see text).
}
\label{fig:confronto-coll-accr-prim}
\end{figure}

As in our previous paper,  the collisions were simulated as impulsive
events  starting at some time $t_{start}$ and  lasting for $\Delta
t_{coll}$ during which time the diffuse gas fraction is set
continuously to some reduced value, including complete removal.    In
Fig.~\ref{fig:confronto-coll-accr-prim} (dot-dash), we present the
case of a single collision, with duration of 130 Myr, starting
arbitrarily at 3 Gyr, using a mass loss history within the  stripping
during the collision obtained numerically by Quilis, Moore, \& Bower
(\cite{quilis00}).  This stripping history removes all diffuse gas
within the galaxy, including that returned by stellar mass loss,
without affecting the molecular clouds directly.  The decrease in the
cloud phase is entirely in response to cloud destruction and
continuing star formation.   We have already tested that the  models
are quite insensitive to the detailed  {\it history} of the event but
depend strongly on the  efficiency of the gas removal in the initial
stage of the collision and on the {\it timing} of the  collision
(VSG05).  In (Fig.~\ref{fig:confronto-coll-accr-prim} a,
dot-dash),\footnote{The figures show only the evolution of the thin disk
component of these systems. Even for the closed standard model,
after about 1 Gyr, all halo star formation has
ceased, that in the thick disk is
substantially reduced, and the thin disk is the only
active zone of the galaxy and
traces the star formation.} the SFR drops after the  removal of the
diffuse gas on the cloud destruction timescale.  The replenishment of
gas occurs only through  processes related to stellar phase, hence
slowly: mass shed by stars that have evolved within the various zones.
No additional mass loss occurs  from the system and the resupply
timescale is  determined by stellar evolution and the assumed
IMF. During the collision, the metallicity initially rises by about
20\% over a very short time, about $\Delta t_{coll}$. Thereafter,  the
SFR drop halts metallicity production for some time, until stellar
evolution resupplies disk gas and heavy elements
(Fig.~\ref{fig:confronto-coll-accr-prim} b, dot-dash).

\begin{figure}
\vspace{5mm}
\centering
\includegraphics[width=8cm]{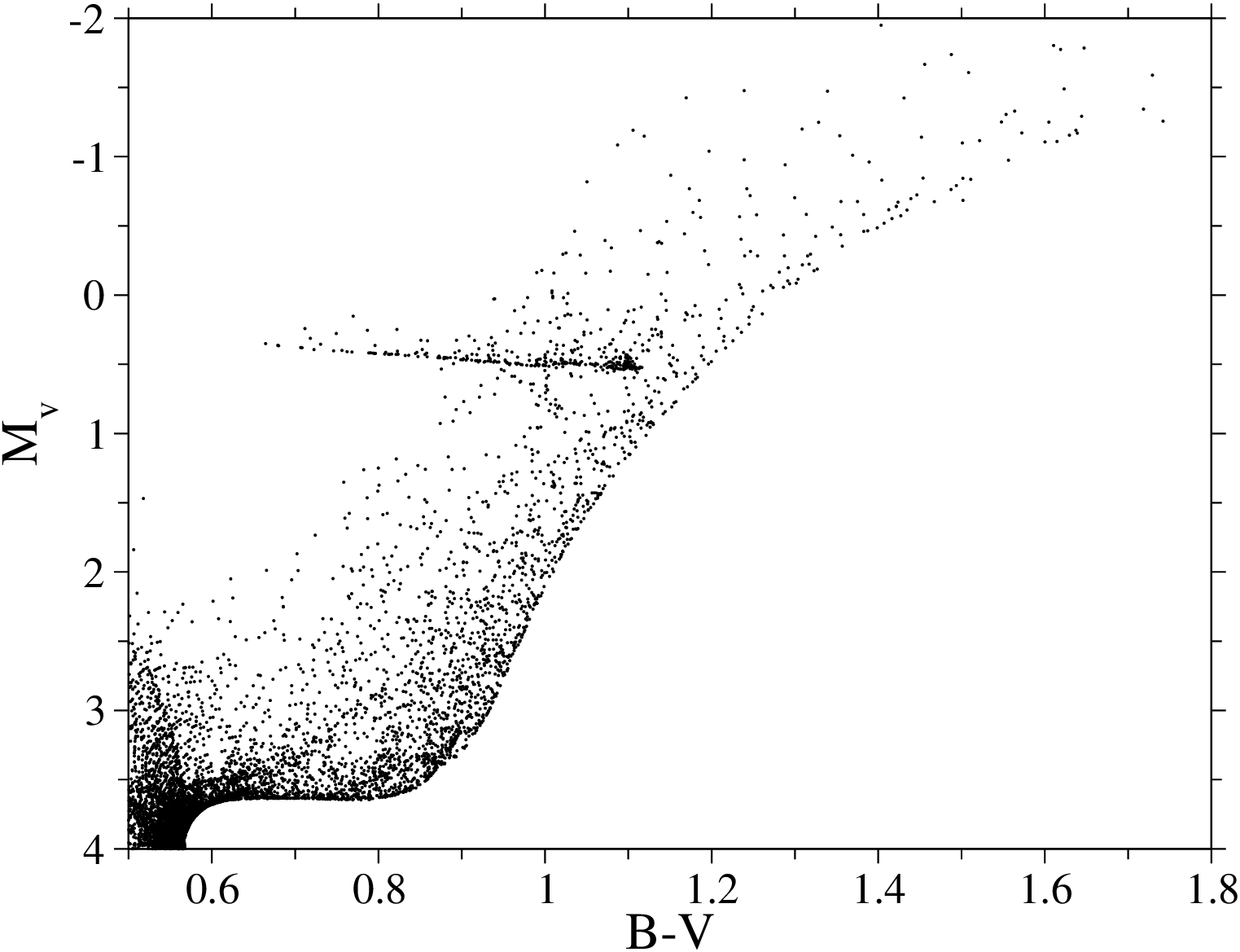}\vspace{15mm}\\
\includegraphics[width=8cm]{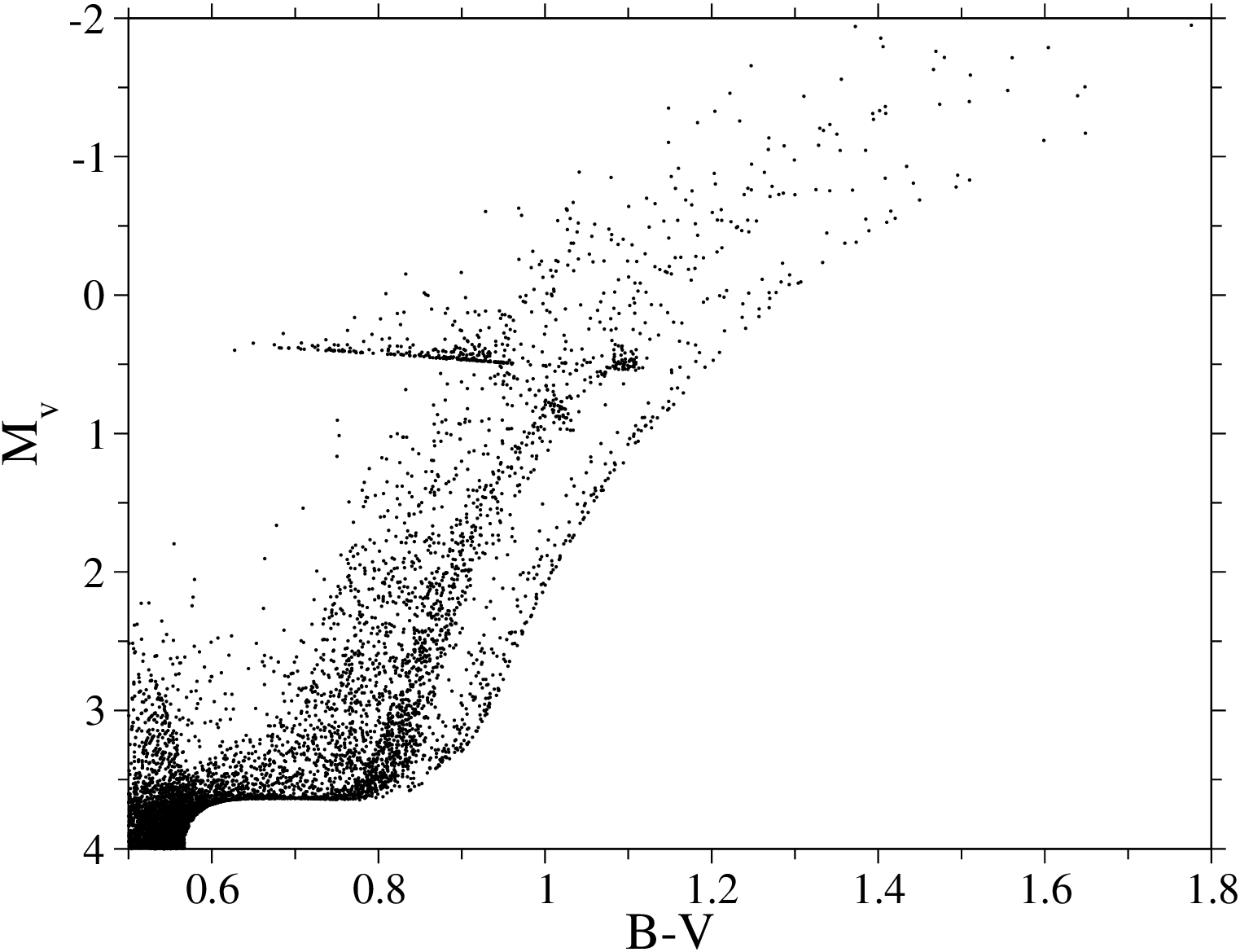}\vspace{15mm}\\
\includegraphics[width=8cm]{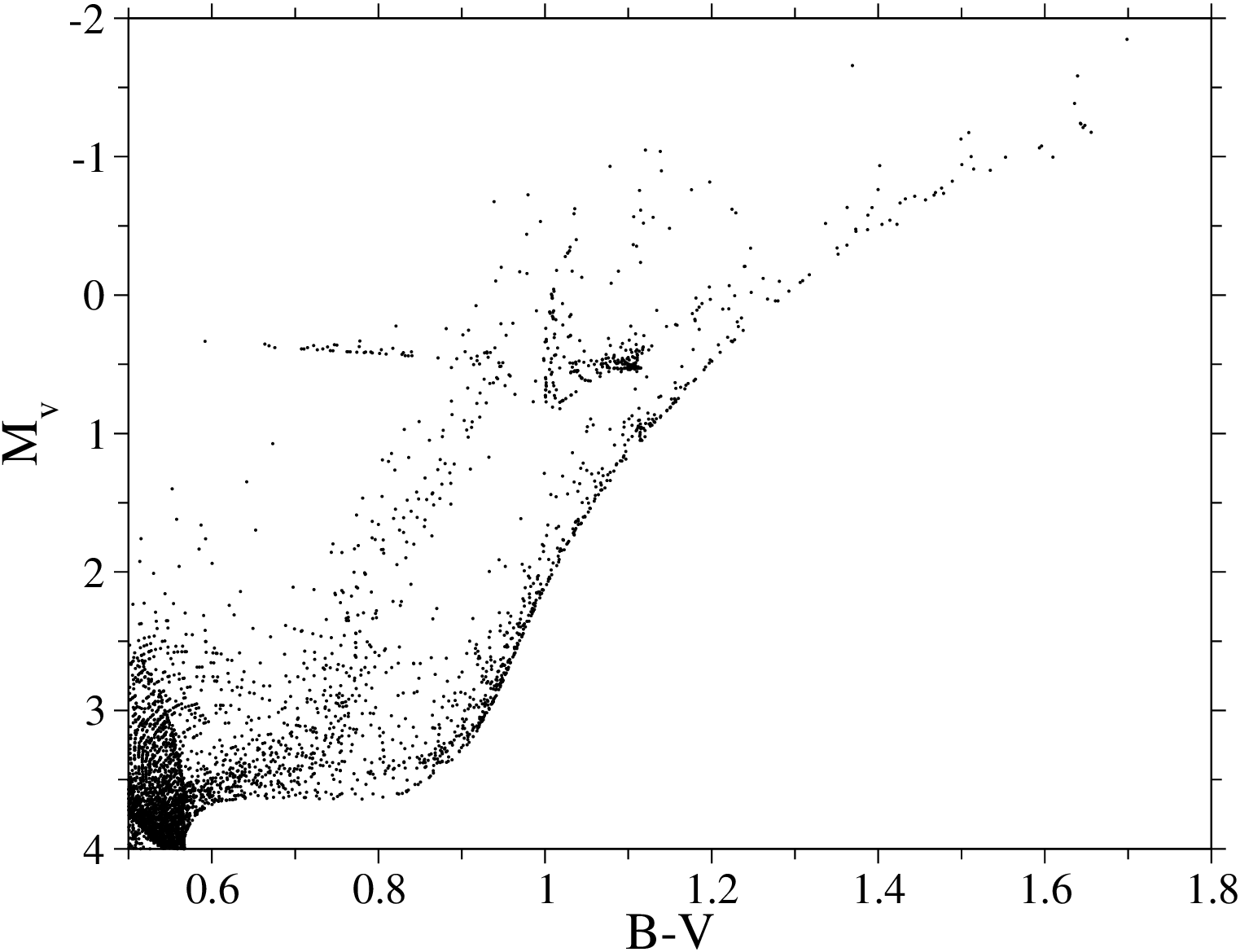}
\caption{Synthetic CM diagram for the cases shown in
Fig.~\ref{fig:confronto-coll-accr-prim}: {\bf (a)} in the standard
case (top), {\bf (b)} in the collision
case (middle) and {\bf (c)} the combined collision-re-accretion
scenario (bottom).  The color-gap in the RGB indicates a
discontinuous change (jump) in the metallicity
(e.g. Fig. 1b, collision case without refilling).  With refilling, the
metallicity produces a broader spread in the two, well-separated, RGBs
with substructure reflecting the dilution due to re-accretion.  For dSph
and Irr, the re-accretion may not occur.}
\label{fig:cmdiagram}
\end{figure}

An important feature for modeling low mass galaxies, in particular,
is that because of the loss of the active phase (the molecular gas)
the SFR after the collision
never fully recovers  its previous levels or those of the standard
model.  Thus, the remnant galaxy (system) permanently stays
metal poor.  As we will discuss in the next section, this behavior is
{\it reminiscent} of the metallicity and star forming histories
of dwarf galaxies in clusters that undergo very
early tidal interactions  while still forming stars.  Notice that in
the CMD (Fig.~\ref{fig:cmdiagram} b), the giant
branch displays a gap with the oldest
population but also a broad  distribution with a mean age now less
than that of the system.  Since we specifically ignore
dynamical mixing within the galaxy and
treat the system as a set of coupled zones, the disk exchanges
matter with the other zones but without rotation.

The final simulation assumes that both collisional stripping  and
accretion of ambient gas can occur.  We showed a wide range of
behaviors for such models in VSG05 citing Vollmer et al. (\cite{Vol01})
who point out the possible role of  re-accretion of stripped gas on
the evolution of cluster galaxies.
Figure~\ref{fig:confronto-coll-accr-prim} (a and b, continuous line)
shows the combined effects of collision and environmental infall of
primordial material.  The dominant effect comes from the removal of
gas  unless the infall rate is extremely high. Accretion doesn't
simply dilute the abundances (Casuso \& Beckman \cite{casuso04}), but
we see instead that -- depending on the metallicity of the accreted
material and the timing and rates  of the stripping and filling events
--  the new gas powers further star formation (VSG05).  The CMD
resulting from this scenario is
shown in Fig.~\ref{fig:cmdiagram} c.  Notice now the
evident separation between the two RGB resulting from the cessation of
star formation at about 2 Gyr.

\section{Discussion}

Multi-population models without dynamics are essentially local and the
results for a larger system represent spatially isolated evolution.  But there
is a simpler type of galaxy for which the model may be more appropriate: dwarf
systems (e.g. Pilyugin \& Ferrini \cite{PF00}).
Dwarf galaxies, particularly because they lack  global patterns such
as density waves, have become the laboratory of choice for studying
the stochastic side of large scale star formation (Mateo \cite{mateo98};
Grebel \cite{greb04}).   Studies of dwarf spheroidals in the Local Group
frequently show complex star forming histories.  For instance, Tolstoy
et al. (\cite{tol03}) summarize the range of behaviors for four dSph
systems.  For Sculptor, they derive an early peak (between 10 and 15
Gyr ago).  For Fornax and Leo I,  in contrast, they find a peak of
activity in the more recent interval, 1 to 8 Gyr, and for Carina they
find multiple episodes -- interpreted as bursts -- throughout the
period less than 10 Gyr ago.  In all cases, however, these galaxies
are now more or less inert.  The metallicity of each system is
significantly below the Galactic disk value, even below the thick
disk.  Similar conclusions were reached by Pritzl et al. (2003) for
the dwarf galaxy HIPASS 01321-37 for which again a peak in the star
formation was found more recently than 10 Gyr.
A recent paper by  Tolstoy et al. (\cite{tol04}) reports evidence for
multiple star forming events in Sculptor.  These, occurring in its
early stages (about 10 Gyr ago) are qualitatively described as a
boom-bust behavior, a series of bursts separated by a hiatus of star
formation.  The authors propose two broad alternative scenarios as
working hypotheses: one in which the star formation is self-suppressed
for some time after an initial burst, the other in which interactions
with the environment -- including other galaxies -- trigger multiple
star forming events and/or produces a cessation of the star formation
for some extended time.  Both behaviors are found in our models, but
we need to be explicit about the qualitative comparison
of behaviors.  We use
a galactic structure including multiple zones.  At least for
Irrs, there is a
large scale structure (halo) including an extended envelope of gas that we
model using a thick disk (which is the main contributor to the evolution
of the system).\footnote{It also appears
possible that the recently studied double red
giant branch in $\omega$ Cen {\it may} be the result of such environmental
action and further hints that this cluster may be the remnant of a
dwarf galaxy (e.g. Sollima et al. \cite{sol05}), but
at this point we do not want to claim too much  for our models.}

\section{Conclusions}

In comparing the resulting
CMDs of interacting systems with those found in the literature, in
particular the just discussed dSph cases, we mean only that the
assumption of a {\it burst} may not be unique; thus
the number of stars formed in any moment and with any
 specific metallicity cannot be formulated in terms of two
simple, independent, global functions $\psi$ and $Z(t)$.  Any non-monotonic
time history -- for instance a cycling, a cessation, or an increase in
the instantaneous star formation rate --  may produce the same basic
result in a coupled system of the kind  we've used.  Even the chemical
evolution alone doesn't remove this ambiguity since the metallicity is
essentially a cumulant of the history of star formation.

\begin{acknowledgements}
We thank the (anonymous) referee for a very
supportive and helpful report, and
John Beckman, Scilla Degl'Innocenti, Pepe Franco,
Joachim K\"oppen, Daniele Galli, Pier Giorgio Prada Moroni,
Jesper Sommer-Larsen, and Eline Tolstoy for preprints,
discussions, and correspondence.  This work was supported by MIUR.
\end{acknowledgements}

\end{document}